\documentclass[aps,pre,10pt,twocolumn,superscriptaddress]{revtex4-2}
\usepackage[utf8]{inputenc}
\usepackage{amsmath}
\usepackage{mathrsfs}
\usepackage{amssymb}
\usepackage{graphicx}
\usepackage{bm}
\usepackage[colorlinks=true,linkcolor=blue,citecolor=blue,urlcolor=blue]{hyperref}
\usepackage{comment}

\newcommand{\apx}[1]{\textsuperscript{#1}}

\begin{document}

\title{Discrete semiclassical theory of quantum-thermal escape}

\author{Luca Salasnich}
\affiliation{\mbox{Dipartimento di Fisica e Astronomia ``Galileo Galilei'', Università di Padova, via Marzolo 8, 35131 Padua, Italy}}
\affiliation{Padua QTech Center, Università di Padova, via Gradenigo 6/A, 35131 Padua, Italy}
\affiliation{Istituto Nazionale di Fisica Nucleare, Sezione di Padova, via Marzolo 8, 35131 Padua, Italy}
\author{Cesare Vianello}
\email{cesare.vianello@phd.unipd.it}
\affiliation{\mbox{Dipartimento di Fisica e Astronomia ``Galileo Galilei'', Università di Padova, via Marzolo 8, 35131 Padua, Italy}}
\affiliation{Istituto Nazionale di Fisica Nucleare, Sezione di Padova, via Marzolo 8, 35131 Padua, Italy}

\begin{abstract}
The dissipationless quantum-thermal escape rate from a metastable well is often written as a Boltzmann average of the Hill-Wheeler flux over a continuum of energies. This continuum treatment diverges exponentially at low temperature, because it pairs the quantum partition function of the well with a classical flux integral. Retaining the discrete nature of the quasibound spectrum removes the divergence exactly, and the resulting rate is obtained in closed form on both sides of the crossover temperature $T_0$. Evaluating a uniform Kemble transmission probability on Bohr-Sommerfeld levels further removes the parabolic-barrier error and recovers the WKB decay rate of the lowest resonance at zero temperature. Benchmarked against the resonances of cubic and quintic potentials, obtained by complex scaling, this uniform semiclassical rate stays within $9\%$ of the exact one over eleven orders of magnitude. Below $T_0$, the rate rises above its zero-temperature plateau with the level spacing as activation energy, and the plateau ends at a temperature $T_1$ noticeably below $T_0$. In an isolated well, whose levels are not replenished, the thermal rate is only the initial slope of a decay which ceases to be single exponential at $T_1$, while at $T_0$ the initial escape moves from tunneling inside the well to the barrier top. We discuss implications for Josephson junctions, single-electron pumps, and nuclear burning in dense stellar matter.
\end{abstract}

\maketitle

\section{Introduction}

Since the pioneering work of Arrhenius \cite{arrhenius}, the escape of a particle from a metastable well is a classical subject with countless applications. It underlies the rate of chemical reactions as much as the decay of a nucleus or of a supercurrent. At low temperature, the passage of the particle through the potential barrier proceeds mainly by quantum tunneling \cite{bell, kastner}, while thermal activation becomes more and more relevant as the temperature increases. At zero temperature, the semiclassical WKB method \cite{wentzel, kramers_wkb, brillouin} leads to the Gamow tunneling factor \cite{gamow,hillwheeler,landau,berry}, while the imaginary part of the analytically continued free energy provides a general description of metastability \cite{langer1,langer2, gunther}. The Euclidean path-integral approach expresses the decay rate in terms of bounce solutions and their fluctuations \cite{coleman,callan}. At finite temperature, the theory of thermally activated escape began with the work of Kramers \cite{kramers} based on classical statistical mechanics, where activation is inseparable from dissipation; the same coupling to a heat bath that supplies the energy to surmount the barrier also damps the motion, and the escape rate depends on the friction coefficient throughout. Most later work has retained that setting; a comprehensive review is given in Ref.~\cite{hanggi}; see also Ref.~\cite{weiss}.

Here we are concerned with the dissipationless case, in which the metastable well is thermally populated but the escape itself is coherent. This is the appropriate description when the coupling to the environment is weak on the scale of the level spacing of the well. It is realized, e.g., in current-biased Josephson junctions of high quality factor \cite{devoret1, devoret2, martinis}, in dynamic quantum dots of single-electron pumps \cite{akmentinsh1, akmentinsh2}, and, through a collective coordinate, in false-vacuum decay in weakly coupled systems and their cold-atom analogues \cite{coleman, callan, rutkevich, mzumdar, lagnese, zenesini}. This is the situation addressed by Affleck \cite{affleck}, who showed that quantum tunneling and thermal activation can then be described within a common semiclassical formalism. His result is compact and widely used: the escape rate is a Boltzmann average of the Hill-Wheeler outgoing flux through a parabolic barrier \cite{hillwheeler} over a continuum of energies. It interpolates between thermal escape at high temperature and quantum tunneling below the crossover temperature $T_0 = \hbar\omega_b/2\pi k_B$, with $\omega_b$ the barrier frequency. It is, however, explicitly a high-temperature construction, expected to hold when the thermal energy exceeds the level spacing of the metastable well, $k_BT > \hbar\omega_a$, with $\omega_a$ the well frequency. The regime where the two energy scales are comparable is not a remote limit. For instance, instanton rate constants for hydrogen-transfer reactions on interstellar ices, computed down to the temperatures of dense molecular
clouds, level off and become essentially independent of temperature \cite{lamberts, meisner}. Such a plateau is the signature of escape from the lowest state of the well, which a continuum treatment of the spectrum fails to reproduce.

The inadequacy of such a continuum expression in the deep-tunneling regime $T<T_0$ manifests in an exponential divergence as $T \to 0$, which originates in pairing the quantum partition function of the well, carrying the zero-point factor $e^{\beta\hbar\omega_a/2}$, with a classical flux integral which does not compensate it \footnote{Not every continuum theory suffers from such a pathology; the low-temperature limit of instanton theory is asymptotic to the WKB rate of tunneling out of the lowest vibrational level \cite{ansari}, the zero-point energy being recovered through the fluctuation prefactor.}. In chemical physics, where the same Boltzmann average over a continuum of energies is the one-dimensional form of the canonical rate of semiclassical transition state theory \cite{miller1, miller2} and of its modern path-integral descendants \cite{richardson, richardson2, dusek}, this spurious factor was identified for unimolecular decomposition in Ref.~\cite{gk}, which concluded that tunneling must be considered from individual quantum states; its signature is the migration of the saddle point of the energy integral below the bottom of the reactant well \cite{fang}. In condensed-matter physics, the correct low-temperature behavior is usually reached within the functional integral approach developed by Grabert, Hänggi, Larkin, Weiss, and others (see Refs.~\cite{hanggi, grabert} and references therein), which reproduces the original results of Caldeira and Leggett \cite{caldeira1, caldeira2, caldeira3}. This however is built around dissipation and reaches the undamped case only as a nontrivial limit. Neither line yields a discrete counterpart of Affleck's formula, whose appeal lies in its simplicity, nor a quantitative statement of what is lost by using the continuum form.

The purpose of this contribution is to supply them, with a set of closed-form evaluations making the size and the origin of the error explicit and a benchmark against numerically exact resonances, and to show where they lead. In Sec.~\ref{sec:theory} we (i) evaluate Affleck’s continuum rate above and below the crossover, exhibiting explicitly the divergence in the zero-temperature limit; (ii) construct the discrete counterpart, obtaining its relation to the continuum formula in each regime and its $T\to0$ limit at the Gamow rate; (iii) remove the independent parabolic-barrier defect by evaluating a Kemble uniform transmission probability \cite{kemble,froman} on Bohr-Sommerfeld levels with the true classical period, so that the zero-temperature limit becomes exactly the WKB decay rate of the lowest resonance; (iv) show that below $T_0$ the rate rises above its plateau with the level spacing as activation energy, and that the plateau ends at a temperature $T_1<T_0$; (v) consider the opposite, isolated limit, in which the well is prepared in thermal equilibrium and then left to itself. Which limit applies depends on whether relaxation within the well is faster or slower than escape. In the isolated limit the levels are not replenished, the probability of remaining in the well is a sum of exponentials, and the thermalized rate is only its initial slope. This gives $T_1$ and $T_0$ a meaning in the time domain: below $T_1$ the decay is single exponential, while at $T_0$ the initial escape moves from tunneling inside the well to the barrier top. In Sec.~\ref{sec:numerics} we test the whole construction against numerically exact resonances obtained by complex scaling \cite{aguilar,balslev,moiseyev,moiseyev_book} in arbitrary-precision arithmetic, and against the exact time evolution of an isolated thermal well. Finally, in Sec.~\ref{sec:conclusions} we summarize the results and discuss their implications for Josephson junctions, isolated systems such as single-electron pumps and false-vacuum decay, and nuclear burning in dense stellar matter.

\section{Discrete semiclassical theory}\label{sec:theory}

We consider a quantum particle of mass $m$ in a one-dimensional potential $U(x)$, having a local minimum (potential well) at $x=x_a$ and a local maximum (barrier top) at $x=x_b$. Let $U_a \equiv U(x_a) = 0$, $U_b\equiv U(x_b) >0$, $\omega_a \equiv \sqrt{U''(x_a)/m}$, and $\omega_b \equiv \sqrt{|U''(x_b)|/m}$. For a narrow quasibound resonance $n$, the outgoing boundary condition gives the complex eigenvalue $\mathcal E_n = E_n - i \hbar\Gamma_n / 2$, where $E_n$ is the resonance position and $\hbar\Gamma_n$ its width, so that $\Gamma_n$ is its decay rate. Assuming the well to be in thermal equilibrium with an environment whose coupling is weak enough to populate the levels $E_n$ without broadening them (dissipationless limit), the quantum-thermal escape rate can be written as
\begin{equation}\label{eq:spectral}
\Gamma = \frac{1}{Z_a}\left[\sum_{n:\,E_n < U_b} \Gamma_n\, e^{-\beta E_n} + \int_{U_b}^\infty {\rm d}E\,\frac{P(E)}{2\pi\hbar}\,e^{-\beta E}\right],
\end{equation}
with $\beta \equiv (k_B T)^{-1}$ the inverse temperature, $Z_a=\sum_{n:\,E_n < U_b} e^{-\beta E_n}$ the partition function of the well, and $P(E)$ the barrier transmission probability. This is a resonance-plus-continuum representation of the thermally averaged escape flux. The sub-barrier spectrum is represented by isolated resonances, and their contributions are weighted by their Boltzmann populations in the metastable well. Above the barrier, resonances overlap and cannot in general be assigned unambiguous individual widths and populations. The escape is thus described collectively in the energy representation by the transmitted flux $P(E)/2\pi\hbar$. The separation at $E=U_b$ is a semiclassical matching prescription: isolated resonances are retained individually below the barrier, while overlapping above-barrier states are treated as a continuum flux. Eq.~\eqref{eq:spectral} describes the thermalized limit, in which the environment refills the levels faster than they decay; we discuss the opposite, isolated limit in Sec.~\ref{sec:isolated}.

\subsection{Affleck's semiclassical formula}

By adopting a semiclassical and harmonic approach, Affleck approximated Eq.~\eqref{eq:spectral} as
\begin{equation}\label{eq:affleck-general}
\Gamma_{\mathrm A} = \frac{1}{Z_a}\int_0^\infty {\rm d}E\,\frac{P_b(E)}{2\pi \hbar}\,e^{-\beta E},
\end{equation}
where $Z_a= [2\sinh(\beta\hbar\omega_a/2)]^{-1}$ is the partition function of a harmonic oscillator of frequency $\omega_a$ and
\begin{equation}\label{eq:phib}
P_b(E) = \frac{1}{1+e^{-2\pi(E-U_b)/\hbar\omega_b}}
\end{equation}
is the Hill-Wheeler transmission probability for a parabolic barrier of height $U_b$ and frequency $\omega_b$ \cite{hillwheeler,landau}. The crossover between thermally assisted quantum tunneling at low temperatures and quantum-corrected thermal hopping at high temperatures is naturally identified by the crossover temperature
\begin{equation}\label{eq:Tcross}
T_0 =\frac{\hbar\omega_b}{2\pi k_B}.
\end{equation}
We will also use the notation $\beta_0 \equiv 1/k_B T_0=2\pi/\hbar\omega_b$.

For $T > T_0$, the dominant energies lie close to the barrier top. Extending the integration range to $-\infty$, one obtains
\begin{equation}\label{eq:highT}
\Gamma_{\mathrm A} \simeq\frac{\omega_b}{2\pi}\frac{\sinh(\beta\hbar\omega_a/2)}{\sin(\beta\hbar\omega_b/2)}\,e^{-\beta U_b}.
\end{equation}
In the classical limit $k_BT \gg \hbar\omega_{a(b)}$, this reduces to 
\begin{equation}
    \Gamma_\mathrm{cl} = \frac{\omega_a}{2\pi}\,e^{-\beta U_b},
\end{equation}
namely the classical Arrhenius escape rate with attempt frequency $\omega_a/2\pi$. The apparent divergence of Eq.~\eqref{eq:highT} as $T\to T_0^+$ is an artifact of having extended the lower limit of integration to $-\infty$; the parent expression \eqref{eq:affleck-general} remains finite there. This is the same spurious divergence that afflicts standard instanton theory and that has recently been addressed in Refs.~\cite{pollak1,pollak2,lawrence} by making use of the Kemble uniform transmission probability. In the language of chemical kinetics, Eq.~\eqref{eq:highT} is the standard result of harmonic transition state theory \cite{eyring, kastner}, where the prefactor of the rate $k_\text{HTST}$ is written as $k_BT/2\pi\hbar$ times the ratio of the quantum harmonic partition functions of the transition structure and of the reactant. In one dimension the reactant has the single mode $\omega_a$ and the transition structure none, the unstable mode being omitted, so that $k_\text{HTST} = (k_B T/2\pi\hbar)(1/Z_a)e^{-\beta U_b}$, with the $Z_a$ of Eq.~\eqref{eq:affleck-general}. This expression accounts for the quantum mechanical zero-point energy but neglects tunneling, which is conventionally restored by multiplying it by the Wigner \cite{wigner} (or Bell \cite{bell1, bell2}) correction $\kappa_\text{W}=(\hbar\omega_b/2k_BT)/\sin(\hbar\omega_b/2k_B T)$. One verifies at once that the product $\kappa_\text{W} k_\text{HTST}$ gives exactly the rate of Eq.~\eqref{eq:highT}. Consistently, the temperature at which $\kappa_\text{W}$ diverges is the crossover temperature $T_0$, and the artificial character of that divergence is what motivates the usual practice of truncating the Wigner correction as $\kappa_W \simeq 1+\frac{1}{24}(\hbar\omega_b/k_BT)^2$ \cite{wigner, kastner}.

For $T < T_0$, the rate is dominated by sub-barrier energies, where $P_b(E) \simeq e^{-\beta_0(U_b-E)}$ is the WKB tunneling probability for a parabolic barrier. The integral is then elementary and yields
\begin{equation}\label{eq:affleck-lowT}
\Gamma_{\rm A}\simeq\frac{\sinh(\beta\hbar\omega_a/2)}{\pi\hbar(\beta-\beta_0)}\,e^{-\beta_0U_b}.
\end{equation}
In the zero-temperature limit, $\Gamma_{\rm A}$ diverges exponentially. This divergence originates from the fact that the factor $Z_a^{-1}=2\sinh(\beta\hbar\omega_a/2)$ carries the zero-point contribution $e^{\beta\hbar\omega_a/2}$ of the well, while the flux integral in Eq.~\eqref{eq:affleck-general} runs from $E=0$ and thus carries no compensating $e^{-\beta E_0}$. If one also takes the classical limit of $Z_a$, namely $Z_a \to Z_a^{\rm cl} = (\beta\hbar\omega_a)^{-1}$, then Eq.~\eqref{eq:affleck-lowT} becomes $\Gamma_{\rm A}^{\rm cl}=(\omega_a/2\pi)[\beta/(\beta-\beta_0)]e^{-\beta_0 U_b}$. In the limit $T\to 0$, it now tends to the finite expression $\Gamma_{\rm A}^{\rm cl} \to (\omega_a/2\pi)e^{-\beta_0 U_b}$, that is the Gamow form \cite{gamow} for an inverted parabolic barrier, evaluated at $E=0$. A fully continuum treatment thus loses the zero-point energy, whereas the quantum-continuum hybrid of Eq.~\eqref{eq:affleck-general} diverges. 

\subsection{Discretized semiclassical formula}

A minimal way to address the problem with Affleck’s result in the low-temperature limit is to keep a harmonic treatment of both the metastable well and the barrier top, while preserving the discrete nature of the quasibound energy spectrum. We therefore consider the discretized expression
\begin{align}\label{eq:discrete}
\Gamma_{\rm D} &= \frac{1}{Z_a}\sum_{n=0}^\infty \frac{\omega_a}{2\pi}\,P_b(E^{\rm ho}_n)\,e^{-\beta E^{\rm ho}_n},
\end{align}
with $E^{\rm ho}_n = \hbar\omega_a(n+\frac{1}{2})$ and the $Z_a$ of Eq.~\eqref{eq:affleck-general}, which is a Boltzmann average of the individual decay rates $\Gamma_n = (\omega_a/2\pi)P_b(E^{\rm ho}_n)$. This relation has a simple flux interpretation. The decay rate of a quasibound state is the outgoing probability flux divided by the probability stored in the metastable well. Semiclassically, a particle with energy $E_n$ encounters the barrier once per classical oscillation period $\tau_a(E_n)$, so the incident flux, for unit probability in the well, is $1/\tau_a(E_n)$. Multiplication by the barrier transmission probability $P(E_n)$ gives
\begin{equation}\label{eq:flux}
\Gamma_n=\frac{P(E_n)}{\tau_a(E_n)}.
\end{equation}
For a harmonic well $\tau_a=2\pi/\omega_a$, while for an inverted harmonic barrier one has the Hill-Wheeler transmission probability $P_b(E)$ given by Eq.~\eqref{eq:phib}.

Above the crossover temperature, where the sum is confined to energies close to the barrier top, Eq.~\eqref{eq:discrete} may be evaluated by Poisson summation, obtaining
\begin{equation}\label{eq:gammaD_high}
    \Gamma_{\rm D} \simeq \frac{\omega_b}{2\pi}\sinh\!\left(\frac{\beta\hbar\omega_a}{2}\right)\!\sum_{m\in \mathbb Z} (-1)^m \frac{e^{-\beta_m U_b}}{\sin(\beta_m \hbar \omega_b/2)},
\end{equation}
where $\beta_m \equiv \beta + 2\pi i m/\hbar\omega_a$. In writing the last equation, the sum has been extended to all $n \in \mathbb Z$, which is the discrete counterpart of extending the integration limit to $-\infty$ to obtain Eq.~\eqref{eq:highT}. Consequently, Eq.~\eqref{eq:gammaD_high} inherits the spurious pole of Eq.~\eqref{eq:highT} at $T=T_0$, while the parent expression \eqref{eq:discrete} remains finite. The $m=0$ term gives in fact Eq.~\eqref{eq:highT}, and the effect of the discreteness of the spectrum is carried by the terms with $m \neq 0$. Because $|\sin(\beta_m\hbar\omega_b/2)|$ grows like $e^{\pi |m|\omega_b/\omega_a}/2$, those terms are exponentially small. Retaining only the $m=\pm 1$ pair, one obtains, in the limit $\sinh(\pi \omega_b/\omega_a)\gg 1$,
\begin{equation}\label{eq:DA_high}
    \frac{\Gamma_{\rm D}}{\Gamma_{\rm A}} \simeq 1 + \frac{2 \sin(\pi\beta/\beta_0)\sin(2\pi U_b/\hbar\omega_a-\pi \beta/\beta_0)}{\sinh(\pi \omega_b/\omega_a)}.
\end{equation}
The difference between the discrete and the continuum formula above the crossover is thus exponentially small in the frequency ratio, of relative order $2/\sinh(\pi\omega_b/\omega_a)$, that is, e.g., $0.17$ for $\omega_a=\omega_b$ and $0.017$ for $\omega_b=\sqrt 3\,\omega_a$. Moreover, the correction vanishes as $1/T$ for $T \to \infty$. This quantifies the expectation that Eq.~\eqref{eq:discrete} should reproduce Eq.~\eqref{eq:affleck-general} at high temperatures, where the continuum limit $\sum_{n=0}^\infty \to \int_0^\infty {\rm d}E/\hbar\omega_a$ applies.

Below the crossover temperature, where $P_b(E_n)$ takes the sub-barrier WKB form, Eq.~\eqref{eq:discrete} reduces to a geometric series, which gives the closed form
\begin{equation}\label{eq:D_lowT}
    \Gamma_{\rm D} \simeq \frac{\omega_a}{2\pi}\frac{\sinh(\beta\hbar\omega_a/2)}{\sinh[(\beta-\beta_0)\hbar\omega_a/2]}\,e^{-\beta_0 U_b}.
\end{equation}
In particular,
\begin{equation}\label{eq:comp_lowT}
    \frac{\Gamma_{\rm A}}{\Gamma_{\rm D}} \simeq \frac{\sinh[(\beta-\beta_0)\hbar\omega_a/2]}{(\beta-\beta_0)\hbar\omega_a/2} \underset{T \to 0}{\to}\infty.
\end{equation}
In the zero-temperature limit, where the ratio of the two hyperbolic sines tends to $e^{\beta_0\hbar\omega_a/2}$, Eq.~\eqref{eq:D_lowT} instead tends to the finite expression
\begin{equation}\label{eq:gamow}
    \Gamma_{\rm D} \underset{T\to 0}{\to} \frac{\omega_a}{2\pi}\,e^{-\beta_0(U_b-E_0)},
\end{equation}
which is the Gamow form for an inverted parabolic barrier, evaluated at the zero-point energy $E_0=\hbar\omega_a/2$. The uncompensated factor $e^{\beta\hbar\omega_a/2}$ responsible for the divergence of Eq.~\eqref{eq:affleck-lowT} is thus canceled term by term by $e^{-\beta E_0}$ in the discretized formula, and the escape rate correctly saturates at the decay rate of the lowest metastable resonance.

\subsection{Uniform semiclassical formula}\label{sec:uniform}

Eq.~\eqref{eq:discrete} repairs the pathological low-temperature behavior of Affleck's formula, but it still carries a second and logically independent drawback of that construction. The Hill-Wheeler flux \eqref{eq:phib} describes an inverted parabola and knows nothing about the form of the barrier away from its top. Therefore it does not reproduce the exact WKB action \cite{landau,berry} for a generic potential, and the limit \eqref{eq:gamow} is the Gamow rate of a parabolic barrier rather than of the actual one. The solution is to adopt the spectral expression \eqref{eq:spectral} with $\Gamma_n$ given by Eq.~\eqref{eq:flux}, but evaluate each ingredient beyond the harmonic approximation. Let
\begin{equation}\label{eq:action}
S(E) \equiv \oint p(x)\,{\rm d}x = 2\int_{x_1(E)}^{x_2(E)}\sqrt{2m[E-U(x)]}\,{\rm d}x
\end{equation}
be the reduced action of the closed orbit in the well, with $x_1<x_2$ the inner turning
points, and let $x_3>x_2$ be the outer turning point beyond the barrier. The
level positions follow from the Bohr-Sommerfeld quantization condition \cite{landau},
\begin{equation}\label{eq:BS}
S(E^{\rm BS}_n)=2\pi\hbar \left(n+\tfrac{1}{2}\right),
\end{equation}
the attempt time is the true classical period
\begin{equation}\label{eq:period}
\tau_a(E)=\frac{{\rm d}S}{{\rm d}E},
\end{equation}
and the transmission probability is given by the uniform Kemble expression \cite{kemble,froman}
\begin{align}\label{eq:kemble}
P(E)&=\frac{1}{1+e^{2\theta(E)/\hbar}},\\
\theta(E) &\equiv \int_{x_2(E)}^{x_3(E)}\sqrt{2m[U(x)-E]}\,{\rm d}x,
\end{align}
with the barrier action analytically continued to negative values above the barrier top. The resulting uniform semiclassical rate is
\begin{align}\label{eq:uniform}
\Gamma_{\rm U} = \frac{1}{Z_a}&\Biggl[\sum_{n:\,E^{\rm BS}_n<U_b}\frac{P(E^{\rm BS}_n)}{\tau_a(E^{\rm BS}_n)}\,e^{-\beta E^{\rm BS}_n}\nonumber\\
&+ \int_{U_b}^\infty {\rm d}E\,\frac{P(E)}{2\pi\hbar}\,e^{-\beta E}\Biggr],
\end{align}
with $Z_a=\sum_{n:\,E_n^{\rm BS}<U_b} \exp(-\beta E^{\rm BS}_n)$. In the zero-temperature limit, this tends to
\begin{equation}
\Gamma_{\rm U} \underset{T \to 0}{\to} \frac{P(E^{\rm BS}_0)}{\tau_a(E^{\rm BS}_0)},
\end{equation}
which is the WKB decay rate of the lowest quasibound state, with the actual barrier action and attempt time.

To disentangle the error due to the harmonic ladder from that due to the parabolic barrier, it is useful to introduce also the intermediate expression
\begin{align}
\Gamma_{\rm HW}=\frac{1}{Z_a}&\Biggl[\sum_{n:\,E_n^{\rm BS}<U_b}\frac{P_b(E^{\rm BS}_n)}{\tau_a(E^{\rm BS}_n)}\,e^{-\beta E^{\rm BS}_n}\nonumber\\
&+ \int_{U_b}^\infty {\rm d}E\,\frac{P_b(E)}{2\pi\hbar}\,e^{-\beta E}\Biggr],
\label{eq:hw}
\end{align}
identical to Eq.~\eqref{eq:uniform} except that the Kemble probability $P(E)$ is replaced by the Hill-Wheeler probability $P_b(E)$ of Eq.~\eqref{eq:phib}. In this way, the two defects of Affleck's formula can be classified independently: passing from a continuum integral to a discrete sum fixes the $T\to0$ divergence, while passing from Hill-Wheeler to Kemble fixes the barrier shape. In addition, passing from the harmonic ladder to Bohr-Sommerfeld quantization fixes the energy levels and attempt frequency. In Sec.~\ref{sec:numerics} we will show how these steps successively improve the agreement with exact numerical results.

The accuracy of semiclassical quantization may be pushed further by retaining the next order in $\hbar$ \cite{dunham, robnik}. The reduced action becomes
\begin{equation}\label{eq:addition1}
    S(E)-\frac{\hbar^2}{24}\frac{\rm d}{{\rm d}E}\oint \frac{U''(x)}{p(x)}\,{\rm d}x,
\end{equation}
from which the Bohr-Sommerfeld levels and the period follow consistently, while the barrier action in the Kemble transmission becomes
\begin{equation}\label{eq:addition2}
    \theta(E) + \frac{\hbar^2}{24}\frac{\rm d}{{\rm d}E}\int_{x_2(E)}^{x_3(E)}\frac{U''(x)}{\sqrt{2m[U(x)-E]}}\,{\rm d}x.
\end{equation}
In practice this costs two additional quadratures. Whether it actually improves the accuracy of the individual decay rates $\Gamma_n$ strongly depends on $n$. Near the barrier top the turning points $x_2$ and $x_3$ approach each other and the classical orbit approaches the separatrix, where the period diverges logarithmically. Each further energy derivative in Eqs.~\eqref{eq:addition1}-\eqref{eq:addition2} then brings an additional power of $\hbar\omega_b/(U_b-E)$. For a level lying a fraction of $\hbar\omega_b$ below the top this parameter is of order one, so the first correction is no longer small. Since the $\hbar$ series is asymptotic, including it can then degrade the decay rate rather than improve it; see Sec.~\ref{sec:numerics}.

The spectral definition \eqref{eq:spectral}, with the concrete meaning given by Eq.~\eqref{eq:uniform}, immediately gives some quantitative statements about the low-temperature behavior of the rate. At temperatures low enough that the second excited level is negligible, to leading order
\begin{equation}\label{eq:lowT}
\frac{\Gamma}{\Gamma_0} \simeq 1+\frac{\Gamma_1}{\Gamma_0}e^{-\beta\Delta},
\end{equation}
where $\Delta \equiv E_1-E_0$ is the energy gap. Thus the excess rate over the zero-temperature plateau is thermally activated with the level spacing. This has to be contrasted with the dissipative case, where in the presence of Ohmic damping $\gamma$ the low-temperature expansion of the rate acquires an additional power-law term due to the thermally excited low-frequency modes of the bath, $\Gamma^{(\gamma)}/\Gamma_0 \propto (\gamma/m\omega_a)(k_B T/\hbar\omega_a)^2$ \cite{grabert-weiss}. Eq.~\eqref{eq:lowT} can therefore be used as a diagnostic of whether the escape is coherent. Moreover, it implies that the ground-state plateau ends around the temperature
\begin{equation}\label{eq:T1}
    T_1 = \frac{\Delta}{k_B \ln(\Gamma_1/\Gamma_0)}.
\end{equation}
We emphasize that $T_1$ may be noticeably below $T_0$. Identifying the temperature at which the rate leaves its plateau with the crossover temperature, and inferring from it the barrier frequency $\omega_b$, would therefore underestimate $\omega_b$.

\subsection{Decay rate for an isolated well}\label{sec:isolated}

Eq.~\eqref{eq:spectral} presupposes that the environment keeps the Boltzmann populations in place while the particle escapes. Consider instead a well prepared in thermal equilibrium and then left to itself. The states above the barrier are not trapped and leave the well within a single transit, in a time of order $1/\omega_a$, much faster than any quasibound state decays. The departure of these states produces an initial drop of the population, exponentially small below $T_0$ but dominant well above it. After this transient, the particle remaining in the well is described by a mixture of the quasibound states with weights $p_n=e^{-\beta E_n}/Z_a$, each decaying with its own rate. Neglecting the short- and late-time deviations from exponential decay of the individual resonances, the survival probability is then
\begin{equation}\label{eq:survival}
    \mathscr P(t) = \sum_{n:\,E_n<U_b} p_n\,e^{-\Gamma_n t}.
\end{equation}
The initial slope and the time integral of $\mathscr P(t)$,
\begin{align}
    \Gamma_{\rm init} &\equiv -\dot {\mathscr P}(0) = \sum_{n:\,E_n<U_b} p_n \Gamma_n,\\
    \tau &\equiv \int_0^\infty \mathscr P(t)\,{\rm d}t = \sum_{n:\,E_n<U_b} \frac{p_n}{\Gamma_n},
\end{align}
are, respectively, the initial escape rate and the mean lifetime of the trapped particle. The effective rate,
\begin{equation}\label{eq:ressum}
    \Gamma_{\rm eff}(t)=-\frac{\dot{\mathscr P}(t)}{\mathscr P(t)} = \frac{\sum_n p_n \Gamma_n\,e^{-\Gamma_n t}}{\sum_n p_n\,e^{-\Gamma_n t}},
\end{equation}
equals $\Gamma_{\rm init}$ at $t=0$ and tends to $\Gamma_0$ for $t \to \infty$. Notice that $\Gamma_{\rm init}$ is exactly the sub-barrier term of Eq.~\eqref{eq:spectral}, therefore the thermalized rate considered in the previous sections is the initial decay rate of an isolated well. The departure from a single exponential can be characterized by the index
\begin{equation}
    R \equiv \Gamma_{\rm init}\,\tau \ge 1,
\end{equation}
which by Cauchy-Schwarz equals one if and only if the decay is single exponential.

The temperature dependence of these quantities can be understood from the semiclassical analysis of Sec.~\ref{sec:uniform}. Below the barrier top, $\Gamma_n\simeq e^{-2\theta(E_n)/\hbar}/\tau_a(E_n)$, so that $p_n\Gamma_n \propto e^{-\beta E_n-2\theta(E_n)/\hbar}$. Higher energy thus costs Boltzmann weight at the rate $\beta$ and gains transmission at the rate $\tau_b(E)/\hbar$, where $\tau_b(E) \equiv -{\rm d}(2\theta)/{\rm d}E$ is the period of the orbit under the barrier. When $\tau_b$ decreases monotonically with energy, the weights $p_n\Gamma_n$ are peaked around the energy $E^*$ satisfying $\tau_b(E^*)=\beta\hbar$. $E^*$ increases with temperature and reaches the value $E^*=U_b$ exactly at $T=T_0$. The weights $p_n/\Gamma_n$ of the mean lifetime instead decrease with energy at all temperatures, so that $\tau \simeq 1/\Gamma_0$ throughout. Therefore $R \simeq \Gamma_{\rm init}/\Gamma_0 \simeq (\Gamma_n/\Gamma_0)|_{E_n \simeq E^*}$ essentially measures how far the dominant energy $E^\ast$ lies above the ground state. Three regimes follow. As long as $E^* < \frac{1}{2}(E_0+E_1)$, i.e., for $T<T_1$, the ground state dominates, $R\simeq1$, and the decay is single exponential. For $T_1 < T < T_0$, the initial escape comes from the levels around $E^\ast$, which empty first, and the decay slows down to the ground-state rate $\Gamma_0$; escape is still tunneling from inside the well, but the decay is no longer single exponential. Above $T_0$ the weights grow all the way to the barrier top, and the initial decay is dominated by the least bound levels and by the states above the barrier, which leave within a single transit. From this perspective, $T_0$ is thus the temperature at which the initial escape ceases to originate from tunneling inside the well and starts to originate from the barrier top; the departure from single-exponential decay sets in at the lower temperature $T_1<T_0$. We test these predictions as well in Sec.~\ref{sec:numerics}.

\begin{table*}[ht]
\caption{Resonance positions and widths of the cubic potential with $U_b=5\,\hbar\omega_a$, in units $\hbar=m=\omega_a=1$. Exact values (4\apx{th} and 5\apx{th} columns) are obtained by complex scaling. $E_n^{\rm ho}$ (2\apx{nd} column) are the harmonic levels, $E_n^{\rm BS}$ (3\apx{rd} column) are the Bohr-Sommerfeld levels. The 6\apx{th} and 7\apx{th} columns differ only through the transmission probability, Kemble $P(E)$ versus Hill-Wheeler $P_b(E)$, and isolate the effect of the barrier shape. The 8\apx{th} column is the fully harmonic approximation, and comparing it with the 7\apx{th} isolates the combined effect of the harmonic levels and harmonic period.}
\label{tab:resonances}
\begin{ruledtabular}
\begin{tabular}{cccccccc}
$n$ & $E_n^{\rm ho}$ & $E_n^{\rm BS}$ & $E_n$ (exact) & $\Gamma_n$ (exact) &
$P(E_n^{\rm BS})/\tau_a(E_n^{\rm BS})$ &
$P_b(E_n^{\rm BS})/\tau_a(E_n^{\rm BS})$ & $(\omega_a/2\pi)P_b(E_n^{\rm ho})$ \\
\hline
0 & 0.5 & 0.4964492 & 0.4946902 & $4.1268\times10^{-15}$ & $3.8775\times10^{-15}$ & $8.0623\times10^{-14}$ & $8.3644\times10^{-14}$ \\
1 & 1.5 & 1.4664374 & 1.4643128 & $7.4573\times10^{-12}$ & $7.3486\times10^{-12}$ & $3.4588\times10^{-11}$ & $4.4790\times10^{-11}$ \\
2 & 2.5 & 2.4013570 & 2.3986448 & $5.6243\times10^{-9}$  & $5.6238\times10^{-9}$  & $1.1809\times10^{-8}$  & $2.3985\times10^{-8}$  \\
3 & 3.5 & 3.2931831 & 3.2893473 & $2.2373\times10^{-6}$  & $2.2657\times10^{-6}$  & $3.0341\times10^{-6}$  & $1.2843\times10^{-5}$  \\
4 & 4.5 & 4.1274397 & 4.1203510 & $4.7055\times10^{-4}$  & $4.8954\times10^{-4}$  & $5.2540\times10^{-4}$  & $6.5928\times10^{-3}$  \\
5 & 5.5 & 4.8650657 & 4.8473493 & $3.4329\times10^{-2}$  & $3.0889\times10^{-2}$  & $3.0923\times10^{-2}$  & $1.5256\times10^{-1}$  \\
\end{tabular}
\end{ruledtabular}
\end{table*}

\section{Numerical tests}\label{sec:numerics}

We test the results of Sec.~\ref{sec:theory} on the potentials 
\begin{equation}\label{eq:family}
    U(x) = m\omega_a^2\left(\frac{x^2}{2}-\lambda \frac{x^p}{p}\right),\qquad p \text{ odd},
\end{equation}
which for every odd $p \ge 3$ rise without bound on the left and have a single escape channel on the right, with $x_b = \lambda^{-1/(p-2)}$, $\omega_b = \sqrt{p-2}\,\omega_a$, and $U_b = {[(p-2)/2p]}m\omega_a^2 x_b^2$. The case $p=3$ is the cubic potential
\begin{equation}\label{eq:cubic}
    U(x) = m\omega_a^2\left(\frac{x^2}{2}-\lambda\frac{x^3}{3}\right),
\end{equation}
a standard model for a current-biased Josephson junction, for nuclear fission and for false-vacuum decay. It has the peculiarity that $\omega_b=\omega_a$. We use units $\hbar=m=\omega_a=1$ and choose $\lambda=1/\sqrt{30}$, namely $U_b=5\, \hbar\omega_a$, so that the well supports six quasibound levels. The exact resonances are obtained by complex scaling \cite{aguilar, balslev, moiseyev, moiseyev_book}, as detailed in Appendix \ref{sec:appendix}, while the Bohr-Sommerfeld levels, periods, and transmission probabilities are obtained by quadratures from Eqs.~\eqref{eq:action}-\eqref{eq:kemble}.

\begin{figure}[t]
    \centering
    \includegraphics[width=\linewidth]{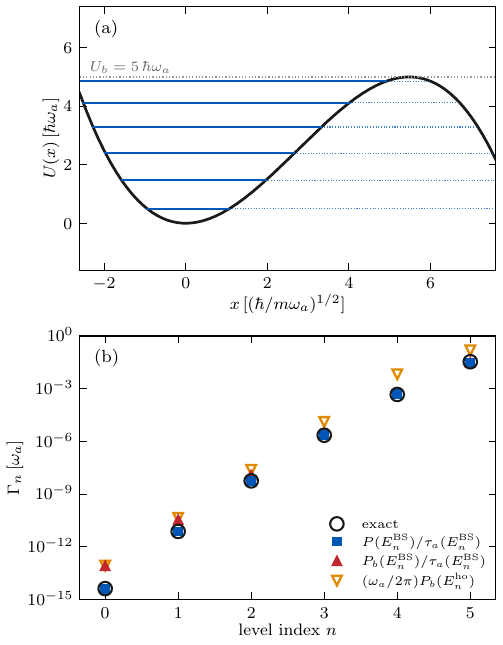}
    \caption{(a) Cubic potential with $U_b=5\,\hbar\omega_a$. Horizontal lines mark the exact quasibound levels $E_n$ of Table \ref{tab:resonances}. (b) Partial decay rates $\Gamma_n$: exact (circles), uniform semiclassical $P(E_n^{\rm BS})/\tau_a(E_n^{\rm BS})$ of Eq.~\eqref{eq:uniform} (squares), Hill-Wheeler $P_b(E_n^{\rm BS})/\tau_a(E_n^{\rm BS})$ of Eq.~\eqref{eq:hw} (filled triangles), and harmonic $(\omega_a/2\pi) P_b(E_n^{\rm ho})$ of Eq.~\eqref{eq:discrete} (open triangles). The uniform values are almost indistinguishable from the exact ones on this scale. The filled triangles approach the exact widths monotonically as the barrier top is reached, showing the effect of the parabolic approximation alone; the open triangles do not, because the harmonic ladder adds an independent error that grows with $n$.}
    \label{fig:1}
\end{figure}

\begin{figure}[t]
    \centering
    \includegraphics[width=\linewidth]{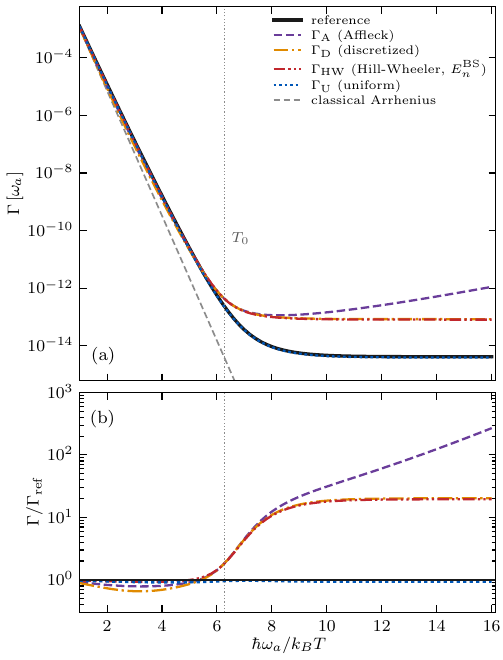}
    \caption{(a) Quantum-thermal escape rate of the cubic potential with $U_b=5\,\hbar\omega_a$ as a function of inverse temperature. Solid black: reference rate \eqref{eq:spectral} obtained from the exact resonances of Table \ref{tab:resonances} and the above-barrier Kemble transmission. Dashed purple: Affleck’s formula \eqref{eq:affleck-general}. Dot-dashed yellow: discretized formula \eqref{eq:discrete}, i.e. the Hill-Wheeler transmission evaluated on the harmonic ladder $E_n^\text{ho}$. Dash-dot-dotted red: intermediate formula \eqref{eq:hw}, i.e. the Hill-Wheeler transmission on Bohr-Sommerfeld levels. Dotted blue: uniform formula \eqref{eq:uniform}, i.e. the Kemble transmission on Bohr-Sommerfeld levels. Comparing the last three curves separates the two sources of error in the discretized formula, namely the harmonic ladder and the parabolic barrier. Dashed grey: classical Arrhenius law. The vertical line marks the crossover temperature $T_0$. (b) Same curves divided by the reference rate.}
    \label{fig:2}
\end{figure}

\subsection{Semiclassical formulas}

Fig.~\ref{fig:1} and Table \ref{tab:resonances} compare the individual decay rates $\Gamma_n$ of the quasibound levels [Eq.~\eqref{eq:flux}]. We first observe that the knowledge of the exact resonances allows us to verify the appropriateness of the energy splitting in Eq.~\eqref{eq:spectral}. The discrete summation over individual rates requires a sharp separation between quasibound states, but the widths grow so rapidly with energy that the topmost states cease to be well defined resonances. The same difficulty is familiar in the theory of magnetic relaxation, where the levels near the top of the anisotropy barrier are the ones that dominate the thermally activated regime and are also the least well defined \cite{garanin}. In the present case, $\Gamma_5 \simeq 0.034\,\omega_a$ is safely smaller than the level spacing, but the next state, at $E_6 \simeq 5.47\,\hbar\omega_a$, has $\Gamma_6 \simeq 0.34\,\omega_a$, comparable to its spacing. Truncating the sum at an arbitrary level would therefore leave the result ambiguous at high temperature. Splitting the rate as in Eq.~\eqref{eq:spectral} avoids this, as the discrete sum is restricted to the sub-barrier levels, which are narrow and unambiguous, while the overlapping states above the barrier are described collectively by the continuum flux $P(E)/2\pi\hbar$.

The uniform semiclassical result entering Eq.~\eqref{eq:uniform} (6\apx{th} column of Table \ref{tab:resonances}) reproduces the exact decay rates to between $0.01\%$ and $10\%$ across thirteen orders of magnitude, the largest deviations occurring for the ground state ($-6\%$), where the WKB approximation is least accurate, and for the highest level ($-10\%$), which lies within $0.15\,\hbar\omega_a$ of the barrier top. We verified that for the ground state, retaining the next order in $\hbar$ in the quantization condition [Eqs.~\eqref{eq:addition1}-\eqref{eq:addition2}] reduces the error of $E^{\rm BS}_0$ from $1.8\times10^{-3}\,\hbar\omega_a$ to $4\times10^{-6}\,\hbar\omega_a$, and the error of the decay rate from $-6\%$ to $+1.1\%$. The same holds for the lowest excited states up to $n=4$, where the decay rates become accurate to better than $0.4\%$. As anticipated, for the topmost level ($n=5$) the trend is reversed. Although its position moves closer to the exact one, its decay rate error deteriorates from $-10\%$ to $-30\%$ due to the divergence of the asymptotic series. 

The Hill-Wheeler expression of Eq.~\eqref{eq:hw} (7\apx{th} column of Table \ref{tab:resonances}), which encodes only the parabolic approximation for the barrier, differs from the exact rates by factors $19.54$, $4.64$, $2.10$, $1.36$, $1.12$, and $0.90$ for $n=0,\dots,5$. The discrepancy decreases monotonically as the level approaches the barrier top, as expected since the parabolic approximation becomes exact at $E=U_b$. Indeed, for $n=5$ the Hill-Wheeler and Kemble values agree to within $0.1\%$ ($0.901$ against $0.900$, as fractions of the exact result), the residual $10\%$ being the ordinary WKB error common to both. 

The harmonic ladder $E_n^{\rm ho}$ used in Eq.~\eqref{eq:discrete} (8\apx{th} column of Table \ref{tab:resonances}) introduces a second and independent error which instead grows with $n$, because anharmonicity pushes the true levels below the harmonic ones by an amount reaching $0.65\,\hbar\omega_a$ at $n=5$. As the tunneling exponent is $\beta_0(U_b-E)$, such a shift is strongly amplified, and indeed $E_5^{\rm ho}=5.5\,\hbar\omega_a$ lies above the barrier top, where $P_b$ has already saturated to unity. The combination of the two errors is therefore non-monotonic in $n$, which explains the irregular behavior of the open markers in Fig.~\ref{fig:1}. The harmonic ladder is instead almost harmless for the ground state; since this is the only resonance that matters in the limit $T \to 0$, the low-temperature plateau of $\Gamma$ will be almost independent of whether one uses $\Gamma_n = (\omega_a/2\pi) P_b(E_n^\text{ho})$ or $\Gamma_n=P_b(E_n^{\rm BS})/\tau_a(E_n^{\rm BS})$. In this limit, the difference with the exact result is entirely due to the parabolic approximation of the barrier.

Fig.~\ref{fig:2} shows the escape rate itself. Above $T_0$, all the semiclassical expressions are in close agreement with each other and with the reference rate (see the discussion below), and follow the classical Arrhenius law up to the quantum prefactor of Eq.~\eqref{eq:highT}. Below $T_0$ the curves separate in the manner anticipated in Sec.~\ref{sec:theory}. Affleck's formula \eqref{eq:affleck-general} turns upward and diverges. At $\hbar\omega_a/k_BT=16$ it already exceeds the exact rate by a factor $269$, and the excess grows as $e^{\beta\hbar\omega_a/2}$, in accordance with Eq.~\eqref{eq:affleck-lowT}. The discretized formula \eqref{eq:discrete} instead saturates to the Gamow form given by Eq.~\eqref{eq:gamow}, which is however 20.3 times too high (this is the ratio $(\omega_a/2\pi)P_b(E^{\rm ho}_0)/\Gamma_0$ of Table \ref{tab:resonances}, since at $T=0$ the rate is the decay rate of the ground state alone). The uniform formula \eqref{eq:uniform} successfully tracks the exact result over the whole temperature range of Fig.~\ref{fig:2}, with a maximum deviation of $8.5\%$ and a $T \to 0$ limit $6\%$ below the exact $\Gamma_0$. Since $\Gamma_{\rm U}$ and $\Gamma$ share the same above-barrier integral, which supplies $89\%$ of the total rate at $\hbar\omega_a/k_B T = 0.5$ and $78\%$ at $\hbar\omega_a/k_B T=1$, their agreement at high temperature is partly by construction, and the significant comparisons are those below $T_0$. 

The intermediate curve $\Gamma_\text{HW}$ shows which of the two approximations is responsible for the difference between $\Gamma_\text{U}$ and $\Gamma_\text{D}$ on the two sides of the crossover. At $\hbar\omega_a/k_B T=1$ one finds $\Gamma_{\rm D}/\Gamma \simeq 0.89$, while $\Gamma_{\rm HW}/\Gamma \simeq 0.98$, essentially equal to $\Gamma_{\rm U}/\Gamma \simeq 0.98$: near and above the crossover the entire error of $\Gamma_{\rm D}$ comes from the harmonic ladder rather than from the parabolic barrier, which is harmless there because the escape occurs close to the top. Indeed, the ratio $\Gamma_{\rm D}/\Gamma_{\rm HW}$ is $0.94$ at $\hbar\omega_a/k_B T = 0.5$, dips to $0.70$ around $\hbar\omega_a/k_B T = 3.2$ and returns to unity below the crossover. The origin of this behavior is the mismatch in the number of sub-barrier levels, which are six for Bohr-Sommerfeld but only five for the harmonic ladder. At high temperature both formulas are dominated by the flux above the barrier, which supplies $89\%$ of the total at $\hbar\omega_a/k_BT=0.5$; there the harmonic ladder is a faithful midpoint quadrature of the continuum integral, the two above-barrier contributions agreeing to within $3\%$, and the ratio stays close to unity. As the temperature is lowered, the sub-barrier sum takes over; being the closest to the barrier top, the missing level has the largest transmission probability and is thus the one that matters the most, producing a marked reduction of $\Gamma_{\rm D}$. At still lower temperatures, the Boltzmann weight concentrates the sum on the lowest levels, where the harmonic and Bohr-Sommerfeld positions are in good agreement, so that the ratio returns close to unity. Indeed, below the crossover temperature the roles of harmonic ladder and parabolic barrier are reversed: at $\hbar\omega_a/k_B T = 16$ one has $\Gamma_{\rm D}/\Gamma \simeq 20.3$, similar to $\Gamma_{\rm HW}/\Gamma \simeq 19.5$, whereas $\Gamma_{\rm U}/\Gamma = 0.94$, so the wrong plateau is due almost entirely to the parabolic barrier. The two defects are thus independent; each dominates in the regime where the other is irrelevant, and only a formula correcting both is accurate throughout. 

\begin{figure}[t]
    \centering
    \includegraphics[width=\linewidth]{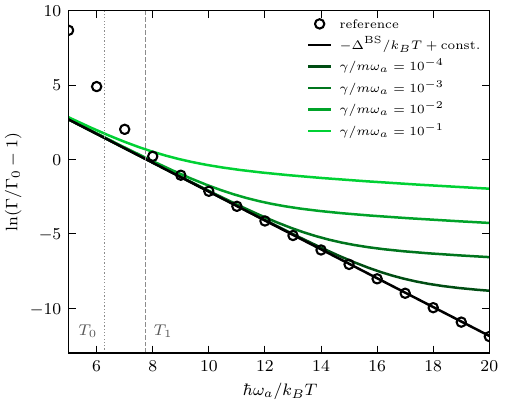}
    \caption{Low-temperature rise of the escape rate above its plateau for the cubic potential with $U_b=5\,\hbar\omega_a$. Each rate is normalized by its own $T\to 0$ limit $\Gamma_0$. Circles: reference rate \eqref{eq:spectral} from the exact resonances of Table \ref{tab:resonances} and the above-barrier Kemble flux. Solid black line: gap law \eqref{eq:lowT} evaluated on the uniform semiclassical levels and rates. Colored lines: gap law plus the Ohmic term $c(\gamma/m\omega_a)(k_B T/\hbar\omega_a)^2$ of Ref.~\cite{grabert-weiss}, with $c=36\pi U_b/\hbar\omega_a$, for $\gamma/m\omega_a=10^{-4}$--$10^{-1}$ (dark to light). Vertical lines mark the crossover temperature $T_0$ and the end-of-plateau temperature $T_1$.}
    \label{fig3}
\end{figure}

Since the cubic potential has $\omega_b=\omega_a$, it cannot probe the distinct roles of the two frequencies. We therefore repeat the whole analysis for the quintic potential ($p=5$), for which $\omega_b = \sqrt{3}\,\omega_a$. We again set $U_b=5\,\hbar\omega_a$, so that the well supports five quasibound levels. The qualitative picture of Fig.~\ref{fig:2} is unchanged, but with some quantitatively different features. First, the crossover scale moves to $k_B T_0 = \hbar\omega_b/2\pi \simeq 0.28\,\hbar\omega_a$, nearly twice the cubic value, and the low-temperature divergence of Affleck's result is faster. At $\hbar\omega_a/k_BT = 16$, that formula now exceeds the exact rate by a factor $724$, against $269$ in the cubic case. This is consistent with Eq.~\eqref{eq:comp_lowT}, since larger $\omega_b$ implies larger $\beta-\beta_0$ at fixed $\beta$. Second, the uniform formula is more accurate, as its maximum deviation from the exact rate over the same range as in Fig.~\ref{fig:2} is $7.8\%$, reached at the low-temperature end of the range. The discretized formula again saturates at a plateau far too high, by a factor $18.4$. The third point is that in the quintic case the potential well is much closer to the harmonic approximation, as the Bohr-Sommerfeld levels are $0.49999$, $1.49913$, $2.49317$, $3.47233$, and $4.41461$ for $n=0,\dots, 4$, deviating from $E_n^{\rm ho}$ by at most $0.09\,\hbar\omega_a$, against $0.65\,\hbar\omega_a$ in the cubic case. Misplacement of the harmonic levels is therefore nearly absent here. Crucially, the two ladders now agree on the number of sub-barrier levels, five in both cases. Accordingly, the difference between $\Gamma_{\rm D}$ and $\Gamma_{\rm HW}$ above the crossover almost disappears; the sub-barrier parts of $\Gamma_{\rm D}$ and $\Gamma_{\rm HW}$ stand in the ratio $1.30$ at $\hbar\omega_a/k_B T=1$ and $1.07$ at $\hbar\omega_a/k_B T=3$, against $0.31$ and $0.58$ for the cubic, and the total ratio $\Gamma_{\rm D}/\Gamma_{\rm HW}$ remains within $8\%$ of unity at all temperatures. The comparison of the two potentials therefore isolates the mechanism rather cleanly, showing that what degrades the accuracy of $\Gamma_{\rm D}$ above the crossover is not so much the displacement of the harmonic levels, but whether that changes how many of them lie below the barrier.

Fig.~\ref{fig3} focuses on the low-temperature behavior of the rate for the cubic potential, showing its rise above the zero-temperature plateau. Below $T_1$ the excess is thermally activated with the gap; evaluated on the uniform semiclassical levels and rates, Eq.~\eqref{eq:lowT} reproduces the exact excess to within $5\%$ for $\hbar\omega_a/k_B T \ge 10$, i.e., $T \lesssim 0.63\,T_0$. Closer to $T_1=0.81\,T_0$ the higher levels start to contribute, and at $\hbar\omega_a/k_BT=8$ the gap law accounts for only $65\%$ of the excess. The linear dependence of $\ln(\Gamma/\Gamma_0-1)$ on $1/k_B T$ well below $T_1$, with negative slope given by the gap energy $\Delta$, is the signature of coherent escape. In the presence of Ohmic dissipation, the excess acquires a power-law term $\propto \gamma T^2$ \cite{grabert-weiss} that overtakes the gap term below a temperature that depends logarithmically on the damping $\gamma$, in our case $0.80\,T_0$ for $\gamma/m\omega_a=10^{-1}$ and $0.38\,T_0$ for $\gamma/m\omega_a=10^{-4}$.

\subsection{Isolated well}

To test Eq.~\eqref{eq:ressum}, we computed the exact time evolution of a particle in the cubic potential prepared in thermal equilibrium in a closed well. The closed potential coincides with $U(x)$ for $x \le x_b$ and is continued beyond the barrier top as $U_b + \frac{1}{2}m\omega_b^2(x-x_b)^2$. Its eigenstates $\chi_m(x)$, with energies $\epsilon_m$, are populated with weights $e^{-\beta \epsilon_m}/Z$. At $t=0$ the closure is removed and each $\chi_m$ evolves under the true Hamiltonian $H$ into $\chi_m(x,t)=[e^{-i H t/\hbar}\chi_m](x)$. The probability of finding the particle in the well is $\mathscr P(t)=Z^{-1}\sum_m \mathscr P_m(t) e^{-\beta\epsilon_m}$, where $\mathscr P_m(t)=\int^{x_b}|\chi_m(x,t)|^2\,{\rm d}x$. The closed-well states above $U_b$ leave within a transit time; afterwards they only rescale $\mathscr P$, and drop out of $\Gamma_{\rm eff}$. The Hamiltonian is discretized on a sinc-DVR grid \cite{lill, muckerman, colbert, light, littlejohn}, with a complex absorbing potential far beyond the barrier, and diagonalized once; the two slowest decay rates are refined in extended precision, and $\dot{\mathscr P}$ is evaluated analytically in the eigenbasis, which preserves its accuracy up to $\omega_a t = 10^{14}$.

\begin{figure}[t]
    \centering
    \includegraphics[width=\linewidth]{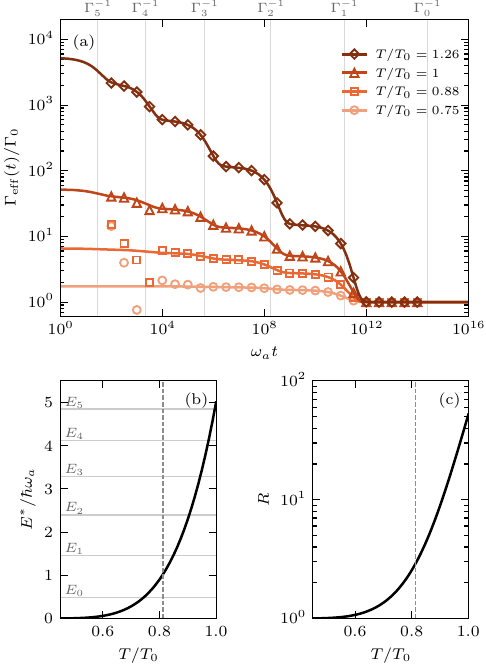}
    \caption{Isolated cubic well with $U_b=5\,\hbar\omega_a$. (a) Effective rate $\Gamma_{\rm eff}(t)/\Gamma_0$ at $T/T_0=0.75$, $0.88$, $1$, and $1.26$ (light to dark). Exact time evolution (open markers) and resonance sum \eqref{eq:ressum} evaluated on the exact resonances of Table \ref{tab:resonances} (solid lines). The uniform semiclassical resonances give curves indistinguishable on this scale. Vertical lines mark the decay times $\Gamma_n^{-1}$ of the quasibound levels. (b) Energy $E^*$ of the dominant contribution to $\Gamma_{\rm init}$ as a function of temperature. (c) Non-exponentiality index $R$ as a function of temperature. Vertical lines in (b) and (c) mark the temperature $T_1$.}
    \label{fig4}
\end{figure}

Fig.~\ref{fig4}(a) compares the resulting effective rate $\Gamma_{\rm eff}(t)=-\dot{\mathscr P}(t)/\mathscr P(t)$ with the resonance sum \eqref{eq:ressum}, evaluated on the exact resonances of Table \ref{tab:resonances}. The two agree to better than $0.5\%$ for $\omega_a t \ge 10^6$ and to better than $0.1\%$ for $\omega_a t \ge 10^7$, at all temperatures. The uniform semiclassical resonances give the same curves to within $6\%$. Since the $\Gamma_n$ differ by two or three orders of magnitude, $\Gamma_{\rm eff}$ is a staircase. For $\Gamma_{n+1}^{-1} \ll t \ll \Gamma_n^{-1}$ the levels above $n$ have emptied, and $\Gamma_{\rm eff}$ has the plateau value $G_n \equiv \sum_{k \le n} p_k\Gamma_k / \sum_{k \le n} p_k$; it thus decreases from $\Gamma_{\rm init}$ to $\Gamma_0$ in steps at $t \simeq \Gamma_n^{-1}$, the last one at $\omega_at \simeq \omega_a\Gamma_1^{-1}=1.3 \times 10^{11}$. The locations of the steps are fixed by $\Gamma_n$, whereas their heights depend on temperature. Specifically, in the step at $\Gamma_n^{-1}$ the rate (in units of $\Gamma_0$) drops by $h_n \equiv (G_n-G_{n-1})/\Gamma_0 \simeq p_n\Gamma_n/p_0\Gamma_0 = (\Gamma_n/\Gamma_0)e^{-\beta(E_n-E_0)}$, where we used $p_0 \simeq 1$ and $\Gamma_n \gg G_{n-1}$. The heights add up to the total drop $\Gamma_{\rm init}/\Gamma_0 \simeq 1+\sum_n h_n$, that is essentially the non-exponentiality index $R$ (Sec.~\ref{sec:isolated}), reported in Fig.~\ref{fig4}(c). As a function of $n$, $h_n$ reflects the competition between the Boltzmann factor and the growth of $\Gamma_n$ that defines the dominant energy $E^*$ (Sec.~\ref{sec:isolated}), shown in Fig.~\ref{fig4}(b): it is largest for the levels near $E^*$ and small for those well above it. At $T=0.75\,T_0$, where $E^*$ is close to $E_0$, $h_n=0.54$, $0.16$, $0.037$, $0.007$, and $0.001$ for $n=1,\dots,5$. The step at $\Gamma_1^{-1}$ carries $72\%$ of the excess $\Gamma_{\rm init}/\Gamma_0-1$, as in the gap law \eqref{eq:lowT}, and those at $\Gamma_4^{-1}$ and $\Gamma_5^{-1}$ less than $0.5\%$ of the rate, hence the staircase is dominated by the step at $\Gamma_1^{-1}$. Raising the temperature enhances the upper steps the most, and they emerge one after the other as $E^*$ climbs through the well. At $T_0$, where $E^*$ reaches the barrier top, $h_n=4.1$, $8.7$, $12.8$, $14.6$, and $11.0$ for $n=1,\dots,5$; all the excited levels contribute comparably, with a maximum at $n=4$ rather than $n=5$ because close to the top the saturation of the transmission and the growth of the period take over. Above $T_0$ the $h_n$ grow up to the topmost level and the earliest steps are the tallest.

At short times ($\omega_a t \lesssim 10^6$), and in particular at low temperature, the exact evolution deviates from Eq.~\eqref{eq:ressum} through quantum beats between the dominant resonance of each initial state and its tiny admixtures of faster resonances (see also Ref.~\cite{janssen}). Expanding $\chi_m(x) = \sum_k c_k^{(m)} \phi_k(x)$ on the outgoing eigenstates $\phi_k(x)$ of $H$ with complex energies $\mathcal E_k = E_k-i\hbar\Gamma_k/2$, which are normalized by the $c$-product $\int \phi_k \phi_l\,{\rm d}x=\delta_{kl}$, so that $c_k^{(m)} = \int \phi_k \chi_m\,{\rm d}x$, we get $\mathscr P_m(t)= \sum_{k,l} c_k^{(m)*}c_l^{(m)} \mathcal O_{kl} e^{i(\mathcal E_k^*-\mathcal E_l)t/\hbar}$, with $\mathcal O_{kl} \equiv \int^{x_b} \phi_k^* \phi_l \,{\rm d}x$. The diagonal terms, $|c_k^{(m)}|^2 \mathcal O_{kk}e^{-\Gamma_kt}$, give the resonance sum \eqref{eq:survival}. The off-diagonal ones survive for two reasons: $c_k^{(m)} \neq \delta_{km}$, because $\chi_m$ is not an eigenstate of $H$, and $\mathcal O_{kl} \neq 0$ for $k \neq l$, because the $\phi_k$ are orthogonal under the $c$-product over all space, not under the Hermitian product restricted to the well. For the ground state $\chi_0$, $|c_k^{(0)}| = 5.8 \times 10^{-11},\,3.1\times 10^{-10},\,1.9 \times 10^{-9},\,1.1\times 10^{-8}$, and $5.3 \times 10^{-8}$, and $|\mathcal O_{0k}| = 8.8 \times 10^{-10},\,7.9 \times 10^{-9},\,5.7\times 10^{-8},\,3.5\times 10^{-7}$, and $1.7\times10^{-6}$, for $k=1,\dots,5$, while $|c_0^{(0)}| \simeq |\mathcal O_{00}| \simeq 1$. To first order in the admixtures, and for $\Gamma_0 t \ll 1$, $\mathscr P_0(t) \simeq 1-\Gamma_0 t + \sum_{k\ge 1}A_k e^{-\Gamma_k t/2}\cos(\omega_{k0}t+\varphi_k)$, with $A_k = 2|c_0^{(0)}c_k^{(0)}\mathcal O_{0k}|$, $\omega_{k0}=(E_k-E_0)/\hbar$, and $\varphi_k$ a phase. Since the amplitudes are extremely small ($A_k=2.2\times10^{-16},\,7.7\times10^{-15}$, and $1.8\times10^{-13}$ for $k=3,\,4,\,5$) the beats are invisible in $\mathscr P_0(t)$; they are bounded oscillations of zero mean, whereas the decay accumulates as $\Gamma_0 t$ and exceeds the largest amplitude, $A_5$, after $\omega_at\simeq45$. In the rate, differentiation multiplies each amplitude by its frequency: $-\dot{\mathscr P}_0(t) \simeq \Gamma_0 + \sum_{k \ge 1} A_k \omega_{k0}e^{-\Gamma_k t/2}\sin(\omega_{k0}t+\varphi_k)$. Since $\Gamma_0 = 4.1 \times 10^{-15}\omega_a$ and $\omega_{k0}$ is of the order of $\omega_a$, the product $A_{k}\omega_{k0}$ can be comparable to, or even exceed, $\Gamma_0$: it is about $190\,\Gamma_0$ for $k=5$, $7\,\Gamma_0$ for $k=4$, and $0.15\,\Gamma_0$ for $k=3$, therefore the beats dominate the effective rate at early times. They change sign with the phase of the oscillation, so that the current can flow back into the well, and persist until $t\simeq2/\Gamma_k$, i.e., $\omega_at\simeq60$ for $k=5$, $4\times10^3$ for $k=4$ and $9\times10^5$ for $k=3$. They matter only at low temperature, where the true rate is itself of order $\Gamma_0$ ($1.7\,\Gamma_0$ at $T=0.75\,T_0$); at $T=T_0$, where the rate is between $26$ and $52\,\Gamma_0$ at these times, the same beats change it by at most $10\%$, and at $T=1.26\,T_0$ they are invisible.

\section{Discussion and Conclusions}\label{sec:conclusions}

We have revisited Affleck's semiclassical formula for the dissipationless quantum-thermal escape rate from a one-dimensional metastable well. Its exponential divergence as $T\to0$ is a failure of the continuum treatment of the quasibound spectrum, as the zero-point factor of the quantum partition function of the well is left uncompensated by the classical flux integral. Keeping the spectrum discrete removes the divergence exactly and at $T=0$ gives the Gamow rate of an inverted parabolic barrier. The residual parabolic-barrier error is an independent defect, removed by using a uniform Kemble transmission evaluated on Bohr-Sommerfeld levels with the true classical period. The resulting formula \eqref{eq:uniform} requires only three one-dimensional quadratures, so it costs no more than Affleck's. For the cubic and quintic wells, it reproduces the exact resonance-based rate to within $9\%$ over eleven orders of magnitude. This includes the zero-temperature plateau, where Affleck's formula fails by an unbounded factor and the discretized one, within the harmonic and parabolic approximations, by a factor of about 20. The individual decay rates are reproduced to within $10\%$. The next order in $\hbar$ brings them to about $1\%$, except for the topmost level, where the asymptotic series breaks down. 

These differences have a practical value, because a barrier height is usually inferred by inverting a measured or computed rate. Below $T_0$ the rate depends on $U_b$ mainly through the tunneling exponent, $\Gamma\propto e^{-\beta_0 U_b}$ for a parabolic barrier. A rate formula off by a factor $F$ therefore biases the barrier by $\delta U_b=k_BT_0\ln F$. For the cubic well this is $-0.2\%$ for the uniform formula ($F=0.94$), $+9.6\%$ for the discretized one ($F=20.3$), and $+17.8\%$ for Affleck’s formula at $\hbar\omega_a/k_BT=16$ ($F=269$). Since $\ln F$ grows as $\beta\hbar\omega_a/2$ for the last of these, the bias increases linearly in $1/T$ without bound. This not only shifts the extracted barrier height, but also distorts its apparent temperature dependence, and would be read as a spurious residual temperature dependence of $U_b$ in a regime where the true rate is already constant. The same arithmetic applies wherever a barrier is inferred from a rate constant in the deep-tunneling regime. For example, for a hydrogen-transfer barrier with $\hbar\omega_b/k_B\simeq1500$~K, i.e., $T_0\simeq240$~K, factors $20$ and $269$ in the rate correspond to $6$ and $11$~kJ\,mol$^{-1}$. These are small against the barrier, but comparable to the accuracy sought from electronic-structure calculations. While the standard analysis of deep tunneling uses instanton theory with the exact bounce action, we argue that the uniform formula \eqref{eq:uniform} is attractive because it is a single closed-form expression covering both the thermal and the tunneling regime.

The semiclassical formula also shows that the rise of the rate above its plateau is thermally activated with the level spacing $\Delta = E_1-E_0$, and that the plateau ends at the temperature $T_1 \propto \Delta$, smaller than the crossover temperature $T_0$. Ohmic damping instead adds a term $\propto \gamma T^2$, so that the excess $\Gamma/\Gamma_0-1$ discriminates coherent from dissipative escape. Current-biased Josephson junctions, the textbook realization of the cubic potential, with $\omega_a=\omega_b=\omega_p$ the plasma frequency, in which macroscopic quantum tunneling and level quantization were first established \cite{devoret2,martinis}, allow a direct test. For the junction of Ref.~\citenum{martinis}, at $u=\Delta U/\hbar\omega_p=2$--$5$, one has $T_0\simeq30$--$36$~mK, and the exact cubic resonances give $T_1\simeq28$--$29$~mK and $\Delta/k_B\simeq170$--$220$~mK. At $25$~mK the escape temperature, defined by $\Gamma=(\omega_p/2\pi)e^{-\Delta U/k_BT_{\rm esc}}$, looks saturated and agrees with the standard undamped value to $1\%$, yet the rate still exceeds $\Gamma_0$ by $25$--$50\%$. With the damping of that junction, $Q\simeq30$, the $T^2$ term is of the same size at $25$~mK and dominates below about $20$~mK. For $Q\gtrsim10^3$ it stays below $1\%$, and between $25$ and $20$~mK the excess should fall from $25$--$50\%$ to $3$--$9\%$ along a straight line of slope $-\Delta$ in $\ln(\Gamma/\Gamma_0-1)$ against $1/T$.

The thermalized rate \eqref{eq:spectral} presupposes that the environment keeps the Boltzmann populations in place. In an isolated well it is only the initial slope of a multi-exponential decay, which relaxes to the ground-state rate $\Gamma_0$ through a staircase of steps at $t\simeq\Gamma_n^{-1}$. The decay ceases to be single exponential at $T_1$, and at $T_0$ the initial escape stops being tunneling from inside the well. The exact time evolution of the cubic well confirms this picture to better than $0.1\%$ at late times, the only deviations being early quantum beats between the resonances admixed in the initial state. This limit applies whenever escape outruns relaxation within the well. In single-electron pumps, the electron is captured in a shallow, effectively cubic dynamic quantum dot, and the limits of fast and slow thermalization, treated numerically in Refs.~\citenum{akmentinsh1,akmentinsh2}, are Eqs.~\eqref{eq:spectral} and \eqref{eq:survival}: at fixed depth they differ only when $R$ departs from unity, that is, above $T_1$. In real-time simulations of thermal false-vacuum decay, the rate prefactor is about eight times below the Euclidean thermal prediction and decreases in time, as the decays deplete the near-barrier modes \cite{pirvu}. The isolated well is the exactly solvable, zero-dimensional quantum counterpart of this depletion: the thermal rate is only $\Gamma_{\rm init}$, and once the upper levels have emptied the rate has dropped by $R$, which reaches $8$ at $0.9\,T_0$ and $52$ at $T_0$ in the cubic well. The exact rates of Sec.~\ref{sec:numerics} also provide a benchmark across $T_0$ for first-principles formulations of finite-temperature tunneling \cite{steingasser2024,steingasser2025}. In cold-atom systems, where false-vacuum decay has been observed \cite{zenesini}, and in one-dimensional traps with tunable barriers, successive $\Gamma_n$ differ by factors $70$--$1800$, so that the steps of $\Gamma_{\rm eff}$ are two to three decades apart in time. Recording the decay of an isolated thermal sample over these decades would measure the individual $\Gamma_n$, and the temperature dependence of the step heights would give the level spacings.

Finally, the burning regimes of dense stellar matter \cite{salpeter,gasques} correspond one to one to the regimes found here: zero-temperature pycnonuclear burning to the ground-state plateau, thermally enhanced pycnonuclear burning to the rise above it, the intermediate regime to the crossover, and thermonuclear burning to activation above $T_0$. Salpeter and Van Horn write the thermally enhanced rate as an explicit Boltzmann sum over the vibrational states of the reacting pair in the lattice, with quanta $0.68$ and $0.52\,k_BT_p$ along and across the bond, $T_p$ being the ion plasma temperature, and then replace it by an integral evaluated by steepest descent \cite{salpeter}, which is the continuum step analyzed in this work. Since their Boltzmann factors are measured from the ground state, the zero-point divergence of Affleck’s formula does not arise. What remains is the discreteness error, which for a parabolic barrier Eq.~\eqref{eq:DA_high} bounds above the crossover by $2/\sinh(\pi\omega_b/\omega_a)$, but which is not controlled when only a few levels contribute. This is the situation of carbon ignition in white-dwarf cores, the prelude to type Ia supernovae, which occurs at $T\sim T_p$, where $k_BT$ spans only one or two vibrational quanta and the rates are uncertain by one to two orders of magnitude \cite{gasques}. The lower boundary of the thermally enhanced regime, $0.5\,T_p/\ln(1/\sqrt\lambda)$ with $\lambda$ the density parameter, has the structure of $T_1$, a gap divided by the logarithm of a rate ratio. Evaluating the discrete sum there, with the screened Coulomb barrier and the anisotropic lattice levels, is a natural extension of this work. Another is weak damping, which interpolates between the thermalized and the isolated limits through the competition between relaxation within the well and escape from it.

\section*{Acknowledgments}
This work is partially supported by the Project ``Frontiere Quantistiche'' (Dipartimenti di Eccellenza) of the Italian Ministry of University and Research and by ``Iniziativa Specifica Quantum'' of INFN.

\appendix
\section{Complex scaling}\label{sec:appendix}

The exact resonances reported in Table \ref{tab:resonances} are obtained by the method of complex scaling \cite{aguilar,balslev,moiseyev, moiseyev_book}. The difficulty this method removes is that a resonance is not an eigenstate of a self-adjoint operator, for the outgoing boundary condition makes the wavenumber complex, $k=k_r-ik_i$ with $k_i>0$. For a potential tending to a constant at large distance, the Gamow function then behaves as $\psi(x)\sim e^{ikx}=e^{ik_rx}e^{k_ix}$ and grows exponentially. The growth is physically meaningful, as the amplitude found far from the well was emitted earlier, when the state was more populated, but it places $\psi$ outside the Hilbert space, so that no expansion in square-integrable functions can converge to it. Complex scaling restores square integrability by rotating the coordinate into the complex plane, $x\to x\,e^{i\phi}$ with $\phi$ real and positive. The transformation is a similarity (not unitary) transformation, and it may therefore change the spectrum. The Hamiltonian becomes
\begin{equation}
H(\phi)=e^{-2i\phi}\frac{p^2}{2m}+U\!\left(x\,e^{i\phi}\right),
\end{equation}
and along the rotated ray the outgoing tail acquires the real exponent $(k_i\cos\phi-k_r\sin\phi) x$, which is negative as soon as $\tan\phi>k_i/k_r$. Therefore the rotation outruns the divergence, the Gamow function becomes normalizable, and the resonance turns into an ordinary discrete eigenvector of $H(\phi)$, accessible by linear algebra. What happens to the remainder of the spectrum is the content of the Aguilar-Balslev-Combes theorem \cite{aguilar,balslev}. Bound states are unchanged, the continuum pivots about each threshold down to $\arg\mathcal{E}=-2\phi$, and a resonance is uncovered as an isolated complex eigenvalue once the rotating continuum has swept past it, after which it ceases to depend on $\phi$. This $\phi$-independence is the practical criterion for identifying physical solutions, since a finite basis inevitably also produces spurious eigenvalues, and those move when $\phi$, the basis size or the basis frequency is varied. 

The potentials \eqref{eq:family} are an extreme case of the picture, because they are unbounded below and support no continuum, so that every point of the spectrum of $H(\phi)$ is a resonance and the selection rests on $\phi$-stability alone. More importantly, the exponential divergence that complex scaling is designed to tame is absent here. Writing $U(x) \simeq -m \omega_a^2\lambda\,x^p/p$ at large $x$, the Gamow function is $\psi(x) \sim x^{-p/4} e^{i A x^{(p+2)/2}}$, where $A$ is a positive constant. The Gamow function is thus square integrable on the real axis for $p \ge 3$, and the role of the rotation is accordingly different. Here there is no analogue of the lower bound $\tan \phi > k_i/k_r$, as an infinitesimal rotation already makes the outgoing solution decay, but there is an upper bound. On the right ray, $\arg x = \phi$, the outgoing solution has modulus $|x|^{-p/4}e^{-A|x|^{(p+2)/2}\sin[(p+2)\phi/2]}$ and is damped for $\phi<2\pi/(p+2)$. On the left ray, $\arg x=\pi+\phi$, where the potential rises, the sine in the previous expression is replaced by a cosine, so the physical solution decays for $\phi < \pi/(p+2)$. This second condition is the stricter one, so the bound is
\begin{equation}\label{eq:theta_bound}
    \phi < \frac{\pi}{p+2} = \begin{cases}
        0.6283 & p=3,\\
        0.4488 & p=5.
    \end{cases}
\end{equation}
In practice we expand $H(\phi)$ in a harmonic-oscillator basis, in which it is banded with bandwidth $p$, and locate the zeros of $\det[H(\phi)-\mathcal{E}]$ by Newton iteration, evaluating the determinant by a banded LU factorization in arbitrary-precision arithmetic. Since the determinant itself underflows, the iteration is run on $h(\mathcal{E})=[d\ln\det(H(\phi)-\mathcal{E})/d\mathcal{E}]^{-1}$, which has a simple zero of unit slope at each resonance. Working at $50$ decimal digits is essential, because the ground-state width is $\sim10^{-15}\omega_a$, fifteen orders of magnitude below the corresponding resonance position, and is entirely lost to round-off in double precision. 

Stability must be assessed separately for the positions and for the widths, as the latter are far more delicate. The resonance positions are stable to 20 significant digits against variations of the basis size ($N=120$-$1200$), of the basis frequency, and of the rotation angle over the whole admissible range. For the quintic, the widths are stable to 12 significant digits within $\phi=0.20$--$0.38$. For the cubic, whose ground-state width is five orders of magnitude smaller relative to the level spacing, the results are stable only within $\phi=0.48$--$0.55$. In fact, the narrower the resonance, the more strongly the outgoing tail must be damped before a harmonic oscillator basis can resolve it, and $\phi$ must accordingly be pushed towards the bound \eqref{eq:theta_bound}. The values quoted in Table \ref{tab:resonances} have been verified against an independent Siegert calculation \cite{siegert, rapedius}, in which the Schrödinger equation is integrated by a Taylor-series marching scheme along a contour running from the real axis on the left to a rotated ray beyond the barrier, with the damped solution imposed at both ends and the eigenvalue located from the vanishing of the Wronskian.

\bibliography{References}

\end{document}